\documentclass[sn-mathphys,Numbered,a4paper]{sn-jnl}

\usepackage{graphicx}%
\usepackage{multirow}%
\usepackage{amsmath,amssymb,amsfonts}%
\usepackage{amsthm}%
\usepackage{mathrsfs}%
\usepackage[title]{appendix}%
\usepackage{xcolor}%
\usepackage{textcomp}%
\usepackage{manyfoot}%
\usepackage{booktabs}%
\usepackage{algorithm}%
\usepackage{algorithmicx}%
\usepackage{algpseudocode}%
\usepackage{listings}%
\usepackage{siunitx}
\usepackage{chngcntr}
\usepackage{caption}
\usepackage{geometry}
\theoremstyle{thmstyleone}%

\theoremstyle{thmstyletwo}%

\theoremstyle{thmstylethree}%

\begin{document}

\title[Article Title]{Nanostructured multiferroic liquids: on the way to fluid ferroelectric magnets}

\author[1]{\fnm{Hajnalka} \sur{N\'adasi}}
\author[2,6]{\fnm{Peter Medle} \sur{Rupnik}}
\author[3]{\fnm{Melvin} \sur{K{\"u}ster}}
\author[1]{\fnm{Alexander} \sur{Jarosik}}

\author[4]{\fnm{Rachel} \sur{Tuffin}}
\author[4]{\fnm{Matthias} \sur{Bremer}}
\author[4]{\fnm{Melanie} \sur{Klasen-Memmer}}

\author[5]{\fnm{Darja} \sur{Lisjak}}
\author[2]{\fnm{Nerea} \sur{Sebasti\'{a}n}}
\author[2]{\fnm{Alenka} \sur{Mertelj}}
\author[3]{\fnm{Frank} \sur{Ludwig}}
\author[1]{\fnm{Alexey} \sur{Eremin}}

\affil[1]{\orgdiv{Institute of Physics}, \orgname{Otto von Guericke University}, \orgaddress{\street{Universit\"atsplatz 2}, \city{Magdeburg}, \postcode{39106},  \country{Germany}}}

\affil[2]{\orgdiv{Jo\v{z}ef Stefan Institute}, \orgaddress{\street{Jamova cesta 39}, \city{Ljubljana}, \postcode{SL-1000},  \country{Slovenia}}}

\affil[3]{\orgdiv{Institute of Electrical Measurement Science and Fundamental Electrical Engineering and Laboratory for Emerging Nanometrology}, \orgname{TU Braunschweig}, \orgaddress{\street{Hans-Sommer-Str. 66}, \city{Braunschweig}, \postcode{38106}, \country{Germany}}}

\affil[4]{\orgdiv{Merck Electronics KGaA}, \orgaddress{\street{Frankfurter Strasse 250}, \city{Darmstadt}, \postcode{64293},  \country{Germany}}}

\affil[5]{\orgdiv{Department for Materials Synthesis, Jo\v{z}ef Stefan Institute}, \orgaddress{\street{Jamova cesta 39}, \city{Ljubljana}, \postcode{SL-1000},  \country{Slovenia}}}

\affil[6]{\orgdiv{Faculty of Mathematics and Physics, University of Ljubljana}, \orgaddress{\street{Jadranska ulica 19}, \city{Ljubljana}, \postcode{SL-1000},  \country{Slovenia}}}

\abstract{Responsiveness to multiple stimuli and adaptivity are paramount for designing smart multifunctional materials.  In soft, partially ordered systems, these features can often be achieved via self-assembly, allowing for the combination of diverse components in a complex nanostructured material. Here, we demonstrate an example of a liquid that simultaneously displays both ferroelectric and ferromagnetic types of order. This material is a nanostructured liquid crystalline hybrid comprising ferrimagnetic barium hexaferrite nanoplatelets suspended in a ferroelectric nematic host. Director-mediated interactions drive the self-assembly of nanoplatelets in an intricate network. Due to the couplings between the polar electric and magnetic types of order,  this material demonstrates magnetically driven electric and nonlinear optical responses, as well as electrically driven magnetic response. Such multiferroic liquids are highly promising for applications in energy harvesting, nonlinear optics, and sensors.
}


\maketitle

\section*{Introduction}\label{sec:intro}

Soft nanostructured materials such as colloids and liquid crystals (LC) have become very interesting for self-assembly behaviour and the ability to control their properties using weak external stimuli~\cite{Li:2012tz}. These materials have a wide range of applications, particularly in display, optical and communication technologies. In addition, they hold significant promise for wearable electronics and healthcare.

\begin{figure*}[t]
\centering
\includegraphics[width=\textwidth]{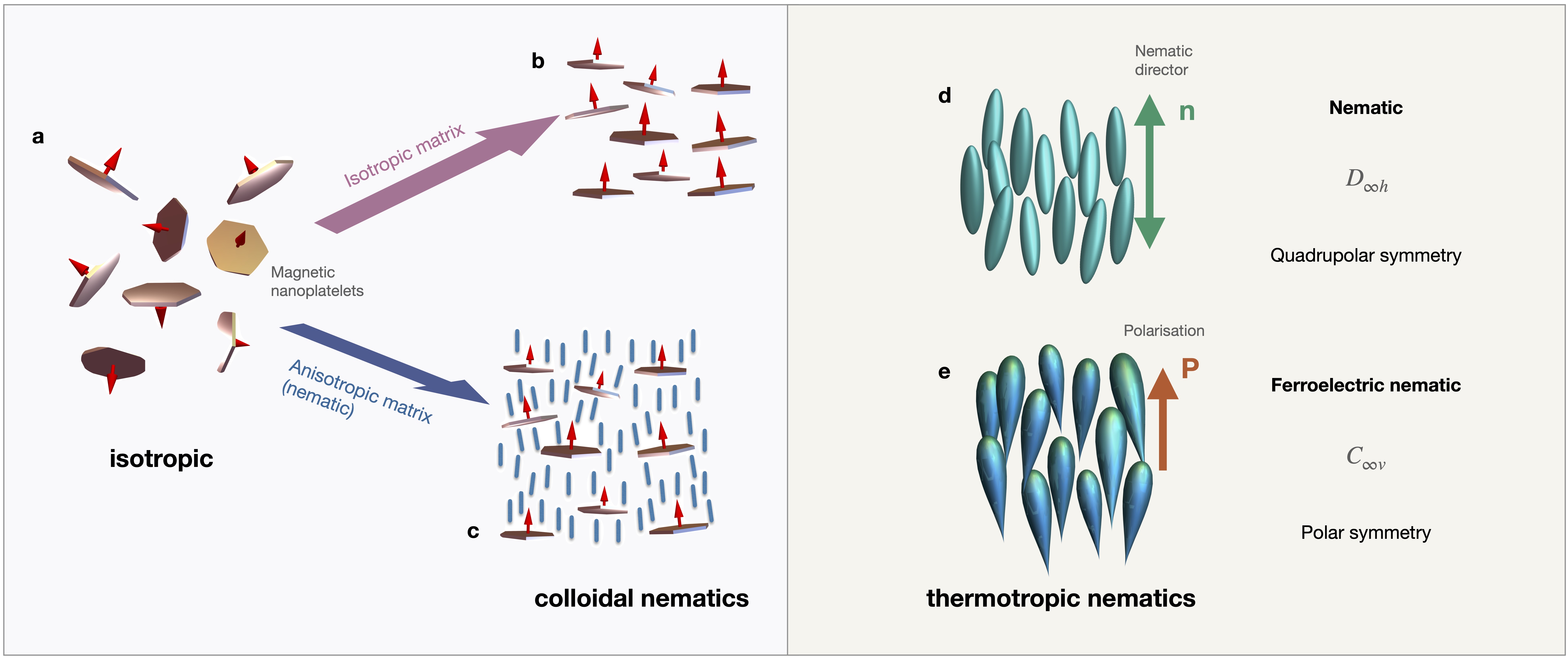}
\caption{\textbf{Fluids with orientational order:} \textbf{a} isotropic suspensions of BaHF nanoplatelets in 1-butanol form \textbf{b} the colloidal ferromagnetic nematic phase at a sufficiently high volume fraction. \textbf{c} Ferromagnetic nematic phase occurs in dispersions of the nanoplatelets in a thermotropic nematic. \textbf{d} Thermotropic nematogens can form the nematic phase N with a cylindric symmetry. \textbf{e} In some compounds, the director invariance is broken, resulting in the ferroelectric nematic phase N$_{\rm{F}}$ with the director's polar symmetry.}
\label{fig:fig1}
\end{figure*}

In solids, intermolecular interactions stabilise ordered crystalline structures, giving rise to materials of various types of order. Some of them can even be ferromagnetic or ferroelectric. In contrast, in liquids, such as liquid crystals and colloids, it is a delicate balance between interactions and entropy accompanied by symmetry-breaking instabilities stabilising different kinds of order~\cite{Chaikin2000, deGennes:2004tf}. Conventional nematics, widely used in modern display technology, have up-down (quadrupolar) symmetry and do not exhibit any spontaneous polarisation.~\cite{deGennes, Oswald:2005wj}. 
Recent years, however, were marked by two important discoveries in the field of liquid crystals: a liquid ferromagnet was discovered in the suspension of scandium-doped barium hexaferrite (BaHF) nanoplatelets (Fig.~\ref{fig:fig1}a-c)~\cite{Sebastian.2018,Shuai:2016fd,Mertelj:2014kv}, and true 3D ferroelectric liquid, a ferroelectric nematic (Fig.~\ref{fig:fig1}d-e), was found to be formed by strongly polar mesogens~\cite{Nishikawa.2017,Mertelj.2018,Sebastian.2020,Mandle.2019co,Mandle.2021,Sebastian.2022}. In contrast to conventional nematics with axial symmetry, these fluids are distinguished by their vector-type properties, such as spontaneous electric polarisation in the ferroelectric nematics N$_{\rm{F}}$, and spontaneous magnetisation in the ferromagnetic nematic N$_{\rm{M}}$. The magnetic structure in N$_{\rm{M}}$ is determined by the self-assembly of the ferrimagnetic nanoplatelets dispersed either in a nematic host or in an isotropic liquid. In the first case, the director-mediated interactions stabilise the order of the nanoplatelets~\cite{Sebastian.2018,Shuai:2016fd,Mertelj:2014kv}. In the latter case, a colloidal nematic is stabilised by the steric, electrostatic and magnetic interactions. Ferroelectric order in nematics was predicted in the early twentieth century but discovered only recently~\cite{Born.2016,Mandle.2017.nematic-nematic,Nishikawa.2017,Chen.2020}. The actual stabilisation mechanism is still under intensive investigation. N$_{\rm{F}}$ materials are distinguished by their unusually high spontaneous polarisation (up to \qty{10}{\micro\coulomb\per\cm^2}), comparable to solid ferroelectrics, and high optical nonlinearity. The latter makes those materials promising for applications in photonics, such as frequency converters and modulators. The fluidity in these hyperpolar materials brings about new extraordinary properties such as super-screening, mechanoelectrical effect, stabilisation of fluid fibres, and electrohydrodynamic instabilities~\cite{Caimi.2023t5j,Sebastian.2023g2e,Rupnik.2024, jarosikPNAS2024, Mathe.2024s4e, Barboza.2022,Marni.2023}. 

Multiferroics are materials that exhibit multiple ferroic behaviors, such as ferroelectric and ferromagnetic properties~\cite{Li.2023,Spaldin.2019,Eerenstein.2006}. Examples of such materials include BiFeO$_3$, EuTiO$_3$, among others. The magnetoelectric properties of multiferroics are crucial for the technology because they enable the manipulation of magnetic properties by an electric field and vice versa. However, the design of multiferroics poses a challenge due to conflicting chemical requirements. Ferroelectricity is favoured by transition metals with empty $d$ orbitals, while ferromagnetism is favoured by partially filled $d$ orbitals~\cite{Spaldin.2019}. Nevertheless, several solutions have been developed to design efficient solid multiferroics, including the use of composite materials~\cite{Zheng.2004, Spaldin.2019}.

In this paper, we demonstrate the first liquid multiferroic composed of ferrimagnetic BaHF nanoplatelets dispersed in a ferroelectric nematic liquid crystal. The suspension forms a hybrid liquid crystal (hybrid-LC) material which remains stable in a wide range of temperatures, including room temperature. In addition to the electro- and magnetooptical responses, this material exhibits direct and converse magnetoelectric responses.

\section*{Main}\label{sec:intro}
\subsection*{Structure and morphology characterisation}

Barium hexaferrite magnetic nanoplatelets have profound effect on the structure and morphology of the multicomponent ferroelectric nematic.
The pure liquid crystal exhibits the high-temperature nematic (N) phase, the intermediate antiferroelectric (M) phase~\cite{Chen.2023}  and the low-temperature ferroelectric nematic (N$_{\rm{F}}$) phase with transitions given below:

\bigskip
\noindent (crystal $<\qty{-20}{\celsius}$) N$_{\rm{F}}$ \qty{45.8}{\celsius} M \qty{57.9}{\celsius} N \qty{87.6}{\celsius}  isotropic\\
\\ 
The nanoplatelets were suspended in the LC host with concentrations ranging from 0.3~wt$\%$ to 4.0~wt$\%$ followed by quenching from the isotropic phase to the N$_{\rm{F}}$ in a magnetic field $\mu_0 H=\qty{482}{\milli\tesla}$. The suspensions did not significantly alter the transition temperatures at this concentration range.

One of the most remarkable features of the nanoplatelet/N$_{\rm{F}}$ hybrids is the multifaceted response to external electric and magnetic fields. This behaviour was characterised in \qty{2.5}{\micro\meter} - \qty{30.0}{\micro\meter} glass planar cells filled with the hybrid liquid crystal. Transparent indium tin oxide (ITO) electrodes on the cell substrates allow applying electric field and measuring the material's response simultaneously. The hybrid material exhibits the ferroelectric-type current transients on the voltage reversal and SHG activity similarly to the pure N$_{\rm{F}}$ phase as well as the residual magnetisation (Supplementary Figs.~\ref{fig:current},~\ref{fig:SIb}).

\begin{figure*}[t]
\centering
\includegraphics[width=\textwidth]{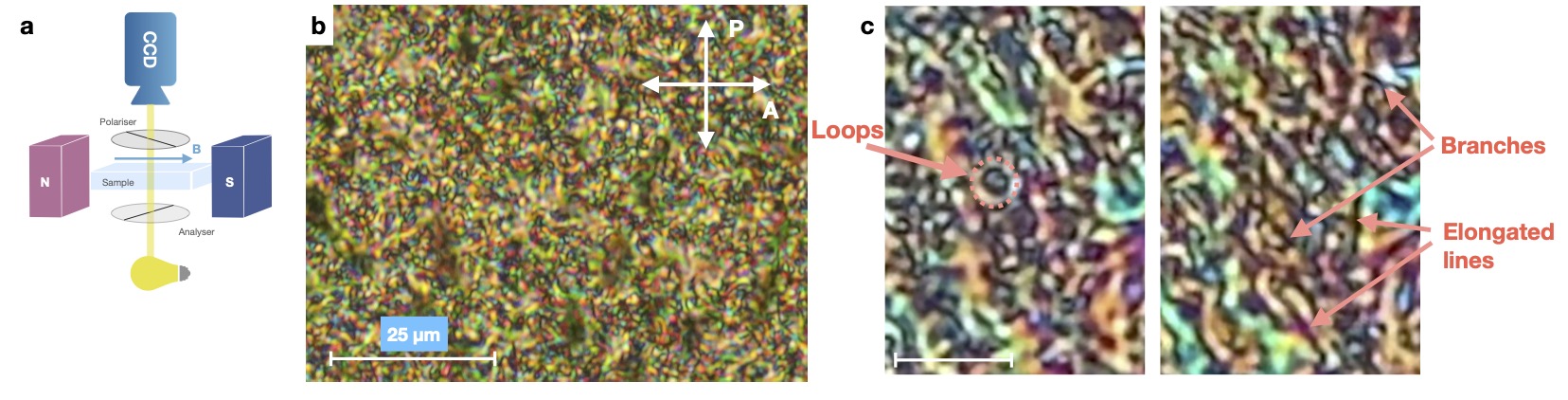}
\caption{\textbf{Optical morphologies in the hybrid-LC:} 
\textbf{a} Schematics of the polarising microscopy setup for optical textures characterisation in a magnetic field.
\textbf{b} Polarising microscopy textures observed in hybrid-LC in \qty{10}{\micro\meter} cell with polyimide rubbed substrates in the N$_{\rm{F}}$ phase at $T=\qty{34}{\celsius}$.  
\textbf{c} Various topologies of disclination lines in a \qty{5}{\micro\meter} cell without aligning layer.  
}\label{fig:textures2}
\end{figure*}

\begin{figure*}[t]
\centering
\includegraphics[width=\textwidth]{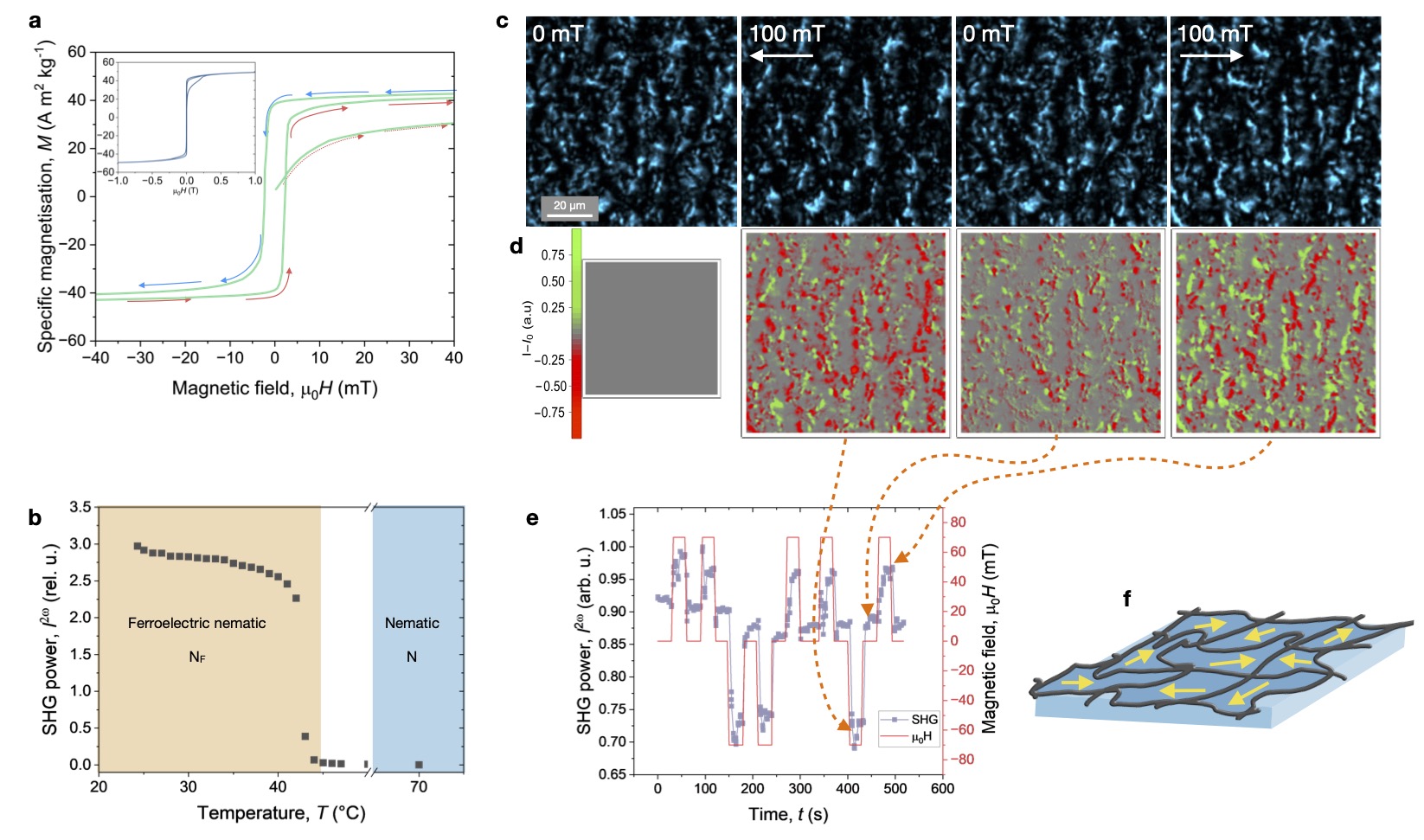}
\caption{\textbf{Magnetic and polar characterisation of the hybrid-LC:}  
\textbf{a} Hysteresis loop of specific magnetization $M(H)$ measured in a 1~wt$\%$ hybrid-LC within a glass capillary, revealing characteristics of ferromagnetic switching. Inset provides the complete field range.
\textbf{b} Variation of spontaneous SHG signal with temperature in a planar-aligned \qty{5}{\micro\meter} cell. 
\textbf{c} Sequence of SHG microscopy images acquired from a \qty{5}{\micro\meter}-thick planar-aligned cell subjected to an external magnetic field (field values indicated on each frame). The full time-lapse is provided in Supplementary Movie 3.
\textbf{d} Corresponding difference images obtained by subtracting the initial field-free frame from each subsequent frame shown in f, highlighting SHG signal changes induced by the field.
\textbf{e} Time plot of the mean SHG intensity, averaged over a \qty{100}{\micro\meter\squared} region of the sample under repeated magnetic field exposure. Orange arrows correspond to selected frames in c/d, marking key time points on the intensity trace.
\textbf{f} A schematic of a polydomain structure of the sample with the arrows indicating the local spontaneous magnetic order. 
}\label{fig:textures3}
\end{figure*}

\begin{figure*}[t]
\centering
\includegraphics[width=\textwidth]{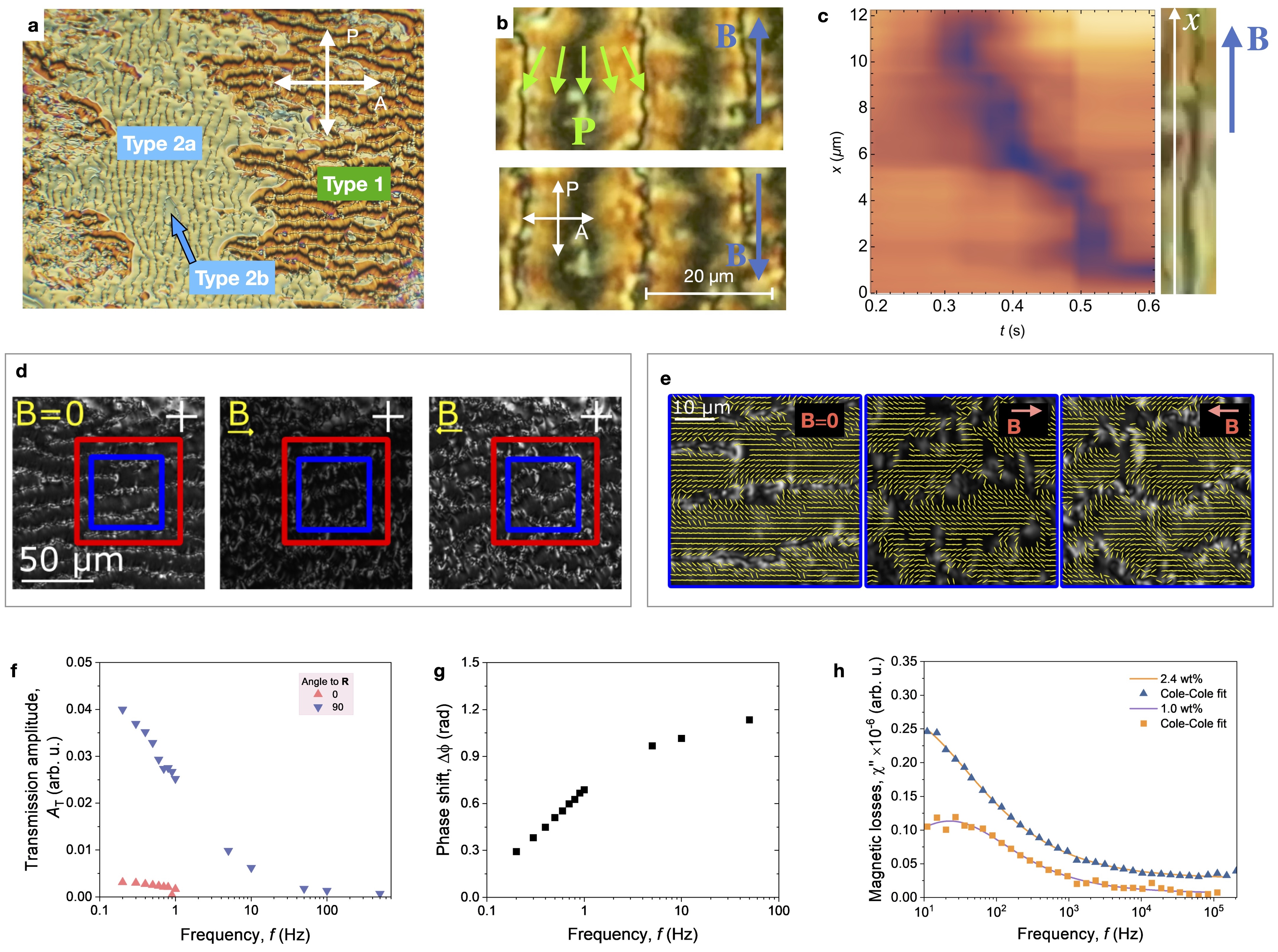}
\caption{\textbf{Magnetooptical behaviour in thin cells} 
\textbf{a} POM image of the N$_{\rm{F}}$-hybrid texture in a \qty{2.5}{\micro\meter} P-type cell between crossed polarisers. 
\textbf{b} Microscopic texture of magnetically prealigned sample in a \qty{2.5}{\micro\meter} thick glass cell with rubbed polyimide aligning layer. The green arrows indicate the spontaneous polarisation as inferred from the SHG characterisation. 
\textbf{c} Space-time plot of the mean transmission measured parallel to a disclination line (displayed on the right) under application of $B\approx\qty{140}{\milli\tesla}$ (Supplementary Movie 6).
\textbf{d} POM images correspond to the cases without a magnetic field and with applied magnetic fields of opposite polarities. 
\textbf{e} the polarisation field extracted from the SHG microscopy marked in yellow is overlain with the POM images marked blue in \textbf{d}. 
\textbf{f} Frequency dependence of the transmission amplitude for the field $\mathbf{B}$ applied parallel (\qty{0}{\degree}) and perpendicular (\qty{90}{\degree}) to the rubbing direction $\mathbf{R}$. 
\textbf{g} Frequency dependence of the phase shift $\Delta\phi$ between the phase of the AC magnetic field and the phase of the induced optical response (transmission) to the field aligned perpendicular to $\mathbf{R}$. (Cell thickness: \qty{3}{\micro\meter}, field amplitude: \qty{5}{\milli\tesla}, measurement area \qty{1}{\mm\squared} (see Supplementary Note 4)) 
\textbf{h} Spectra of magnetic losses $\chi''(f)$ of nanoplatelet/N$_{\rm{F}}$ hybrid recorded at room temperature for two nanoplatelet concentrations, 1.0~wt$\%$ and 2.4~wt$\%$.}\label{fig:thincells}
\end{figure*}

Optical polarising microscopy (POM, Fig.~\ref{fig:textures2}a) provides a tool to explore the molecular order and alignment on micro- and macroscopic scales (Supplementary Note 3 and Figs.~\ref{fig:SItextures},~\ref{fig:SItextures2}). In the N$_{\rm{F}}$ phase, the morphology of the hybrid material, as revealed by POM, primarily depends on the cell thickness and the surface treatment (Figs.~\ref{fig:textures2}a-c, \ref{fig:thincells}a,b, Supplementary Note 4 and Figs.~\ref{fig:SIp1} and \ref{fig:SIp2}). 
Cells with rubbed polyimide treatment and surfactant cetyltrimethylammonium bromide surface layers were used for planar (P-cells) and vertical (V-cells) alignments in the nematic, respectively. 

In both cell types, irregular textures strongly depend on the filling conditions. Diverse morphologies were found in different regions across the same cell, as described in Supplementary Note 4 and Figs.~\ref{fig:SIp1} and \ref{fig:SIp2}. Interestingly, in some areas of V-cells, a response to the magnetic field was observed in small uniform patches (Supplementary Fig.~\ref{fig:SIp1}d, e).

In thick ($d\geq\qty{5}{\micro\meter}$) cells, the most common birefringent optical textures exhibited a disordered, grainy morphology responsive to electric and magnetic fields (Fig.~\ref{fig:textures2}b, Supplementary Fig.~\ref{fig:SItextures}, Supplementary Movie 1, 2). Close inspection of the POM textures revealed a dissordered grainy texture, indicating that the structure of this phase consists of a complex network of entangled lines (Fig.~\ref{fig:textures3}a). Although the precise structure of these lines remains unresolved at this stage, they are commonly interpreted as topological line defects (disclinations) or inversion walls, characteristic of nematic and ferroelectric nematic phases. A clearer picture of the lines is obtained in thin cells, where the single lines can be well resolved. This morphology was observed in thick layers of all studied suspensions, irrespective of the anchoring conditions.

The textural changes in response to external fields are particularly evident in cells with a thickness of \qty{5}{\micro\meter}. Despite being highly disordered, the network of threads readily rearranged in response to a magnetic field $\mathbf{B}$ (Supplementary Movie 2).
The material's ferromagnetic nature became evident upon repeated stimulation by  $\mathbf{B}$. When the field was removed, the established texture only partially relaxed.
Reapplying a field with the same polarity caused only minor changes in the morphology. However, when a field of the opposite polarity was applied, the texture underwent complete reassembly, indicating that the ground state of the hybrid N$_{\rm{F}}$ phase was ferromagnetic (Supplementary Movies 1).
The remanent magnetisation of the cells was demonstrated by direct measurements of the magnetisation using a fluxgate probe (Supplementary Fig.~\ref{fig:SIb}). 
The measurements of the magnetisation in thin \qty{1}{\milli\meter} capillaries using SQUID magnetometer confirmed hysteresis behaviour of the magnetisation (Fig.~\ref{fig:textures2}d). The specific magnetisation per unit mass of the magnetic component in the hybrid reaches saturation at approx.~\qty{40}{\ampere\meter\squared\per\kilogram}, which is in good agreement with the values obtained from powder samples of BaHF platelets~\cite{Lisjak2023}.

Two experimental techniques allow probing the polar structure of mesophases: measurements of the current transients on polarisation reversal and detection of the nonlinear optical second harmonic generation (SHG). Noncentrosymmetric structures of a ferroelectric phase with $C_{\infty v}$ symmetry generate non-zero components of the second-order hyperpolarisability tensor allowing for SHG. In the phases with a centrosymmetric structure, such as the nonpolar nematic phase, SHG is forbidden.

 In the non-doped LC, the ferroelectric N$_{\rm{F}}$ phase has large spontaneous polarisation aligned along the nematic director~\cite{jarosikPNAS2024}. The hybrid-LC shows a similar SHG activity in the temperature range of the N$_{\rm{F}}$ phase.
The SHG signal $I^{2\omega}(T)$ of a sample cell slightly decreases as the temperature increases, as shown in Fig.~\ref{fig:textures3}b. Only when the transition to the M phase occurs does the signal drop to the background level, indicating that the M phase is not SHG active. To gain a better understanding of the polar behaviour, we utilised SHG microscopy.

Figures~\ref{fig:textures3}c, d display the SHG microscopy images of a \qty{5}{\micro\meter} thick nanoplatelet/N$_{\rm{F}}$ cell. The field-free state showcases a very disordered texture, consisting of grains with different orientations of the polar domains (Figs.~\ref{fig:textures3}c, f). 
In a magnetic field,  $I^{2\omega}$ increases, indicating local alignment of the polar axes. Similarly to the optical textures, removing the magnets causes partial relaxation. However, if the field is repeatedly applied with the same polarity, the response diminishes, indicating that the magnetic state of the sample is prealigned.
Reversing the magnetic field causes a significant reorganisation of the network, leading to changes in the $I^{2\omega}$ signal (Fig.~\ref{fig:textures3}e). This nonlinear optical behaviour reflects the coupling between magnetisation and electric polarisation, demonstrating the magnetoelectric properties of the hybrid material.

To suppress the formation of the disordered entangled state, we confined the liquid crystal in thin \qty{2.5}{\micro\meter} cells. The POM images show large domains on a millimetre scale (Fig.~\ref{fig:thincells}a, Supplementary Fig.~\ref{fig:SItextures2}a, b).
The domains are distinguished by their optical transmittance and their behaviour upon the sample rotation or insertion of a wave plate retarder. 
Some domains (Type 1) display high contrast between states of transmission maxima and minima upon a 90$^{\circ}$ rotation between crossed polarisers. 
In contrast, Type 2a and 2b domains show no extinction, suggesting a twisted director structure with opposite handednesses (Supplementary Note 3, Fig.~\ref{fig:SItextures2}c). 
Type 1 domains exhibit a regular pattern of thin and thick lines with a quasi-periodic structure on a scale of \qty{20}{\micro\meter}  at room temperature (Fig.~\ref{fig:thincells}b). The optical transmission analysis suggests a splay deformation of the director with the alignment along the striped pattern in the middle between the thin disclination lines, corresponding to the $\pi$-flip of the polar director. (Supplementary Figs.~\ref{fig:SItextures2}d-g,~\ref{fig:SIp3}).
The stripes respond to an external magnetic field (Supplementary Movie 4). The preferred direction is determined by the magnetic field applied during sample annealing. When a magnetic field is applied along this direction, the pattern undergoes a minor rearrangement, resulting in the straightening of the disclination lines (Fig.~\ref{fig:thincells}b top). A field in the opposite direction disturbs the disclination lines leading to a buckling instability (Fig.~\ref{fig:thincells}b, bottom). These observations imply that the nanoplatelets are incorporated into the disclination lines, with a component of their magnetisation aligned parallel to the lines. The particle's coupling to the nematic director in the volume occurs via the director anchoring at the disclinations' interface. 
Additionally, the particles are anchored in the disclination lines and cannot freely rotate under moderate fields. A threshold field in the range of \qty{120}{\milli\tesla} is required for reassembly of the lines. Such switching occurs in a soliton-like fashion as a propagating deformation  (Supplementary Movie 5).

As shown in Fig.~\ref{fig:thincells}c, propagating director distortions were observed upon the application of a magnetic field of approximately \qty{140}{\milli\tesla}. Significant textural transformations occur upon field inversion, suggesting bistability in the system. In contrast, applying a magnetic field of the same polarity induces only minor distortions. However, strong anchoring of the director at the substrates restricts reorientation away from the disclination lines.
 
 The polar structure of the striped pattern revealed by SHG microscopy indicates the highest SHG efficiency for the polarisation direction primarily along the stripes allowing to map the local polarisation field (Fig.~\ref{fig:thincells}d,e, Supplementary Note 5). In agreement with the polarising microscopy, the director exhibits a splay deformation with an oblique anchoring at the defect lines. 
Realignment of the defect lines by the external magnetic field deforms the director field by bending the director via anchoring thoroughgoingly at the defect lines. However, in such thin cells, most lines are firmly attached to the substrate and cannot be easily reconfigured. In this case, applying magnetic field reduces the SH signal for the primary beam polarised along the stripes.

To characterise the optical response in magnetically aligned samples, we used oscillating magnetic fields applied parallel and perpendicular to the alignment direction $\mathbf{R}$. The samples were confined in a \qty{3}{\micro\meter} cells with parallel rubbing, and a twist-free region was selected for the measurement. The maximal field amplitude was 5~mT, which is much smaller than the field required for the magnetisation reversal. Fig.~\ref{fig:thincells}f, g show the optical transmittance recorded between crossed polarisers. Only small transmittance change was observed when the field was applied along the alignment direction $\mathbf{R}$. Application of the field perpendicular to $\mathbf{R}$ results in a response following the driving field. Since the sample does not have a perfect uniform alignment, but consists of distributed polydomains, the transmission is significantly higher in the field-free state.

\subsubsection*{Magnetically-induced electric effect}
In nematics, magnetic field allows manipulation of the effective birefringence and the orientation of the optical axis. 
However, the magnetic field-director coupling does not allow converting magnetic field into  electric voltage. 
\begin{figure*}[t]
\centering
\includegraphics[width=\textwidth]{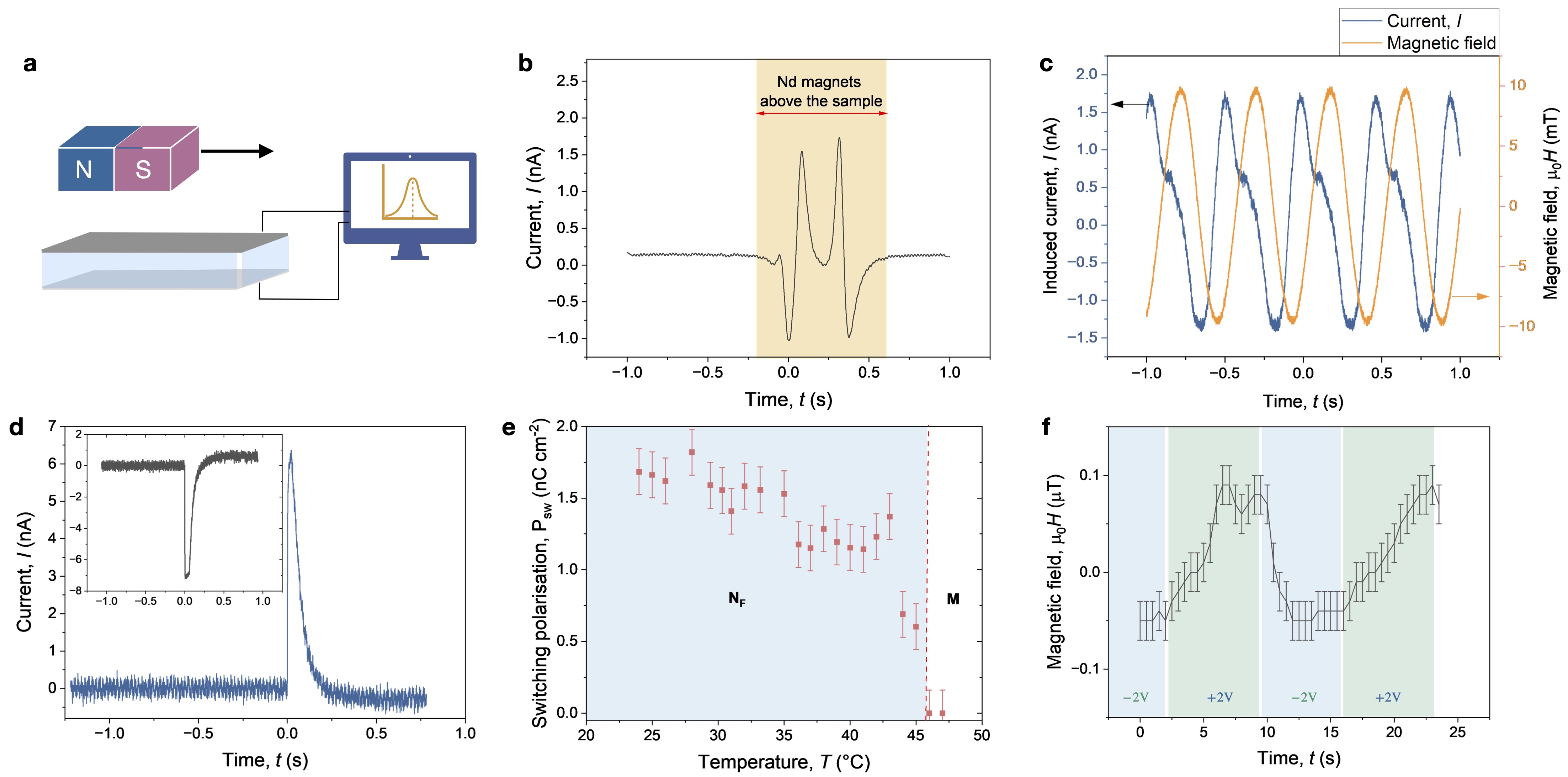}
\caption{\textbf{Magnetically-induced electric and electrically-induced magnetic effects:} \textbf{a} Schematic demonstration of the magnetoelectric effect: a neodymium (NdFeB) magnet passing over a sample cell induces an electric current shown in \textbf{b}. \textbf{c} Electric current induced in a cell placed on a magnetic stirrer which generates a rotating magnetic field (amplitude \qty{10}{\milli\tesla}, \qty{2.0}{\hertz}).  \textbf{d} Current transient measured in magnetically aligned samples upon inversion of the magnetic field (\qty{500}{\milli\tesla}) in a \qty{10}{\micro\meter} thick cell. The inset shows the response in the field of opposite polarity. \textbf{e} Switching polarisation calculated by integration of the current transient. \textbf{f} Electrically-induced magnetic response on poling the sample cell using \qty{2}{\volt} square-wave pulses in samples with 4 wt$\%$ nanoplatelet at room temperature measured by a fluxgate sensor on top of the cell.}\label{fig:response1}
\end{figure*}
In multiferroic nematic, however, the electric polarisation induced by the magnetic field drives the current transient in the connected circuit. To directly demonstrate this effect, a tiny permanent magnet was hovered over a \qty{5}{\micro\meter} thin layer of N$_{\rm{F}}$-hybrid that lies between two ITO electrodes, as illustrated in Fig.~\ref{fig:response1}a,b. The electric current generated, as shown in Fig.~\ref{fig:response1}b, exemplifies the fundamental principle of a basic magnetic sensor based on hybrid N$_{\rm{F}}$.

Another demonstration is given in Fig.~\ref{fig:response1}c, displaying the current response of an LC cell placed on a heat plate with a magnetic stirrer producing a rotating magnetic field. The magnetic field generates current at the same frequency, showing a complex but periodic behaviour. This effect is observed only in the temperature range of the multiferroic nematic phase, and it disappears upon the transition into the high-temperature M phase.

 In these experiments, however, the magnetic field is not uniform. Therefore to explore the magnetically-induced electric response, we placed the LC cell in a uniform magnetic field of an electromagnet allowing to produce magnetic fields up to $\mu_0 H=\qty{650}{\milli\tesla}$. The magnetically-induced electric response to switching the magnetic field on from $\qty{0}{\milli\tesla}$ to $\qty{500}{\milli\tesla}$ is shown in Fig.~\ref{fig:response1}d. Integration of the current transient allows us determining the switching polarisation $P_{\mathrm{sw}}$ whose temperature dependence is given in Fig.~\ref{fig:response1}e. The polarisation is nearly temperature-independent in the temperature range of the N$_{\rm{F}}$ phase and abruptly decreases upon the transition into the M phase. The switching has a distinct polar character: changing the  polarity of the magnetic field results in the inversion of the current response sign and, consequently, the sign of $P_{\mathrm{sw}}$ (inset in Fig.~\ref{fig:response1}d).

\subsubsection*{Converse magnetoelectric effect }
The multiferroic character of the material assumes a converse magnetoelectric effect, where an electric field applied to the LC induces a magnetic response. To test this effect, we placed a sensitive fluxgate sensor on top of a \qty{30}{\micro\meter} cell with sandwiched planar electrodes. As the samples were prepared by quenching in a magnetic field, the sample cells have a small remanent magnetisation as shown in Supplementary Fig.~\ref{fig:SIb}, where the field is recorded. This magnetisation can be detected by inserting the cell under the fluxgate detector, measuring the magnetic flux density at the top of the sample. Flipping the cell results in the inversion of the magnetic field sign, suggesting that the cell acts as a small fluid magnet.

Application of a DC voltage (\qty{2}{\volt}) results in a magnetic response shown in Fig.~\ref{fig:response1}f. This response is very small but still detectable. Voltage inversion results in a reduction of the magnetisation. However, the induced magnetic field decreases with time due to the loss of alignment, particle diffusion and electrically-induced flows in the cell.

\section*{Discussion and outlook}\label{sec12}
The behaviour of the composite multiferroic nanoplatelet/N$_{\rm{F}}$ phase differs significantly from that of the separate components' ferroelectric or ferromagnetic nematic phases. 
 Instead, the volume of the material is filled with a network of threadlike disclination lines (Fig.~\ref{fig:textures2}b,  c).
\begin{figure*}[t]
\centering
\includegraphics[width=0.8\textwidth]{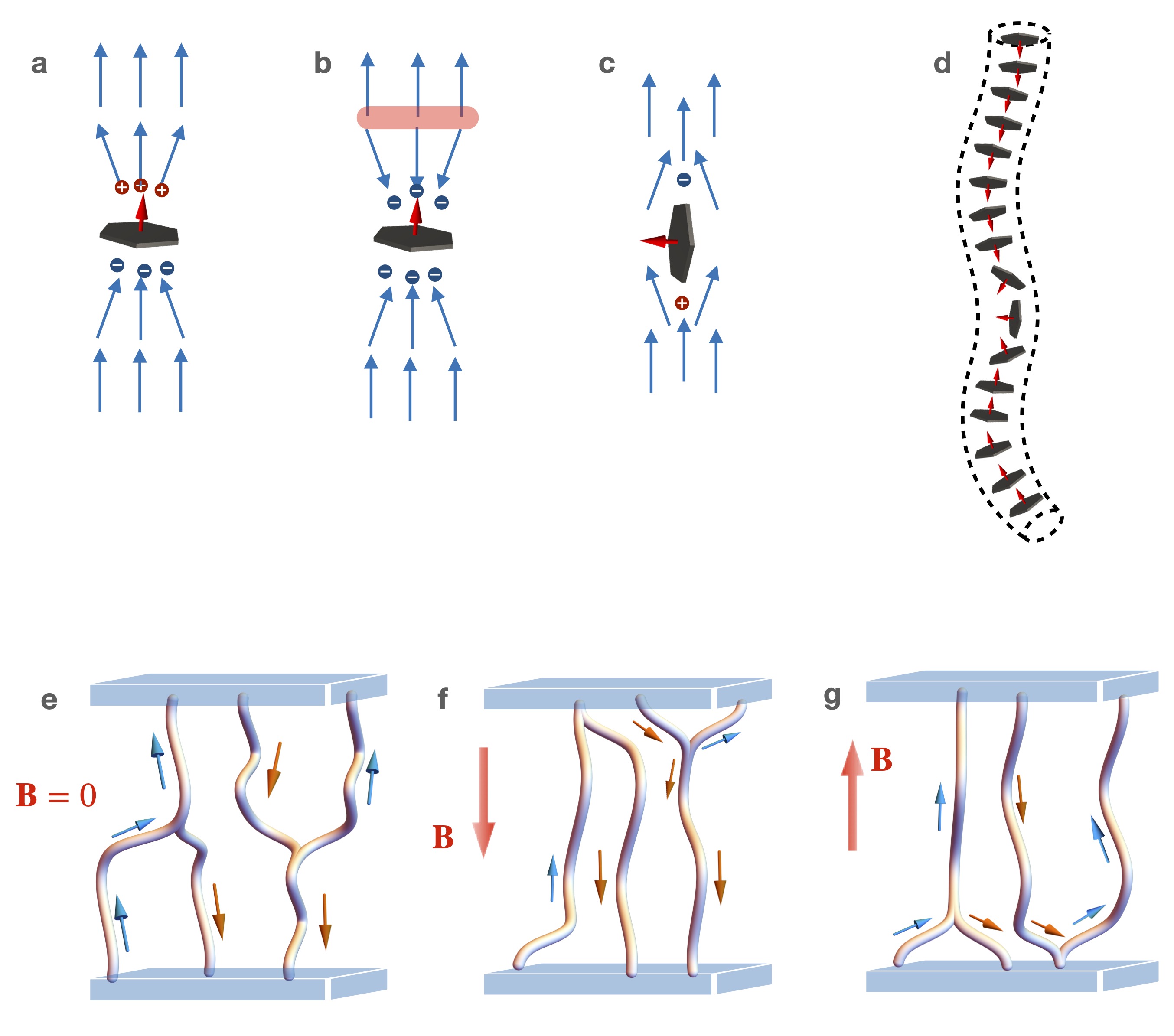}
\caption{\textbf{Proposed structures and network arrangements in the multiferroic N$_{\rm{F}}$ state:} 
\textbf{a} Orthogonal anchoring of the polar director (blue arrow) 
with continuous normal component at the nanoplatelet's surface creating opposite polarisation charges. 
\textbf{b} Discontinuous orthogonal anchoring of the polar director requires an inversion wall (marked as a pink line) to comply with the uniform director in the far field.
\textbf{c} Planar anchoring of the director resulting in the topological (and electric) dipole.
\textbf{d} A model schematics of a disclination line with self-assembled nanoparticles forming a magnetic inversion wall. \textbf{e}-\textbf{g} Magnetisation switching scheme with an example of tripled disclination lines. The blue and orange arrows show the magnetisation direction. \textbf{e} In the field-free state the disclination link in the middle provides a state without magnetisation. \textbf{f} and \textbf{g} Depending on the polarity of the applied magnetic field $\mathbf{B}$ magnetisation is established by the displacement of the link.}\label{fig:schemes}
\end{figure*}

The observed behaviour of lines in response to an external magnetic field indicates that magnetic nanoparticles are confined within them (Fig.~\ref{fig:schemes}). Meanwhile, the occasional occurrence of uniform magnetically responsive domains (Supplementary Fig.~\ref{fig:SIp2}d, e) suggests that some nanoplatelets are dispersed throughout the bulk material.

The dynamics of magnetic nanoparticles are affected by the strong coupling to the director and the constraints of the nematic. Even in oscillating magnetic fields as small as \qty{5}{\milli\tesla}, frequency-dependent optical response can be observed (Fig.~\ref{fig:thincells}f, g). The magnetic dynamics exhibit slow relaxation in the ACS spectra (Fig.~\ref{fig:thincells}h). The relaxation rates can be estimated from fitting the AC susceptometry spectra with a Cole-Cole  model:
\begin{equation}
  \chi(f)=\chi_{\infty}+\frac{\Delta \chi}{1+\left(2 \pi\mathrm{i}  f \tau_{\mathrm{B}}\right)^{1-\alpha}} .
\end{equation}
\noindent where $\chi_{\infty}$ is the susceptibility in the high-frequency limit, $\Delta \chi$ is the relaxation strength of the mode and $\tau_{\mathrm{B}}$ is the  Brownian relaxation time constant. 
Compared to the isotropic suspension of nanoplatelets~\cite{Nadasi.20237to}, the relaxation rate is strongly reduced by more than two orders of magnitude from \qty{700}{\hertz} to \qty{27}{\hertz}, suggesting strong coupling between the LC and the nanoplatelets.

Low compatibility of the nanoplatelets with the bulk N$_{\rm{F}}$ can be explained by the frustration imposed by the polar symmetry of the director and the anchoring conditions. 
Uniform director alignment in ferromagnetic nematics necessitates orthogonal anchoring of the polar director at the nanoplatelet surface, which can be either asymmetric (Fig.\ref{fig:schemes}a) or symmetric (Fig.\ref{fig:schemes}b). Asymmetric anchoring preserves the continuity of the polar director across the nanoplatelet, maintaining a uniform orientation above and below. In contrast, symmetric orthogonal anchoring disrupts this symmetry, requiring the formation of complex defect structures—specifically, director inversion walls—to satisfy the uniform far-field alignment (Fig.~\ref{fig:schemes}b). In the latter scenario, substantial electrostatic energy penalties arise due to the emergence of localised splay domains, which induce polarisation splay and the generation of bound charges
$\rho_{\mathrm{b}}=-\nabla\cdot \mathbf{P}$. 
The planar anchoring would reduce the distortion energy costs, leaving two charged defects and two topological charges at the opposite sides of the platelet (Fig.~\ref{fig:schemes}c). In both cases, a and b, the particles have a dipolar property, favouring the self-assembly of particles in dipolar chains.
 
Defect lines with isotropic order within the cores may favour 1D magnetised structures of nanoplatelets like those proposed on the sketch in Fig.~\ref{fig:schemes}d. Additional charge at the nanoplatelets' surfaces would stabilise oblique anchoring of the nematic director at the defect lines.

In the proposed scenarios, the nanoplatelets themselves can initiate the formation of the disclination lines and their self-assembly.  
Also the high affinity of nanoparticles to the disclination lines can be responsible for the formation of the magnetic nanochains which turn into entangled 3D networks with loops and knots. 

The external magnetic field brings about the restructuring of the network observed in the experiment. The coupling between the magnetisation and the nematic director controls the polar state, resulting in the magnetoelectric effect. To sustain the remanent magnetisation, the topology of the network should be compatible with the vectorial symmetry. Since the orientation of the lines is given by the 1D magnetic order of the nanoplatelets, the remanent magnetisation will be sustained for the lines pinned at the substrate or locked by the neighbouring knots as suggested in the scheme Fig.~\ref{fig:schemes}e-g.

The switching occurs via magnetic inversion within the disclination lines (Fig.~\ref{fig:schemes}a) or displacement of bifurcation points (Fig.~\ref{fig:schemes}b-d). The inversion can be realised through the propagation of the soliton-like structures along the disclination lines. Such propagating fronts can be seen as disturbances in the disclinations observed in thin cells in polarising microscopy.

 In case of the looped knots, inversion walls of the magnetisation within the disclinations can stabilise the magnetised state. Branched structures with mobile links attached to the inversion walls can determine the switching mechanism, allowing the magnetisation to switch between the two magnetised states (Fig.~\ref{fig:schemes}e-g).
This explains why exposure to magnetic fields ($> \qty{0.1}{\tesla}$) is required to obtain the magnetised state. The magnetically mediated (viscous) mechanoelectric effects are responsible for the electric response because it has been shown that the viscous mechanoelectric response exists in the pure ferroelectric nematic system~\cite{Rupnik.2024}. The magnetised domain walls/disclinations that separate the ferroelectric domains rearrange, leading to a transient magnetoelectric response. Although the material is locally ferroelectric, as demonstrated by nonlinear optics, understanding the mechanism of the polarisation-magnetisation coupling on the global scale (in bulk) remains challenging. Studies of colloidal knots in nematics showed that they generate topologically non-trivial structures of topological defects such as boojums, resulting in elastic couplings between the particles and the defects~\cite{Martinez:2014gn}. The director, confined in such a region, is fixed by the topological constraints and can be controlled via alignment of the defects. Thus, the reorganisation of the links in the network allows the realignment of the polarised domains within the network in a polar fashion.

The high degree of disorder in the magnetic network leads to the reduction of the electrically induced magnetic response compared to the magnetically-induced electric response. The results shown here open a promising route for the development of soft multiferroic systems. The coupling between the magnetic and electric order types can be improved by either tailored functionalisation of the nanoplatelets or designing asymmetric Janus-type platelets.

These room-temperature soft multiferroics, exhibiting coupled magnetic and electric order, hold promise not only for advancing fundamental science but also for their strong applicative potential, creating opportunities for soft and flexible sensors and multi-stimuli responsive actuators in soft robotics and wearable technology.

\section*{Methods}\label{sec11}
The multiferroic suspension was prepared by mixing 0.9 wt$\%$ BaHF ferrofluid with M5 ferroelectric liquid crystal mixture (Merck) to achieve a final BaHF concentration ranging from 0.3 to 4 w$\%$. The mixture was heated to \qty{100}{\degree} to fully evaporate the carrier fluid 1-butanol and quenched on top of neodymium magnets (\qty{482}{\milli\tesla}) to room temperature. The BaHF suspension was prepared as described in \cite{Bostjancic.2019u5p,Nadasi.20237to}.

 To characterise the optical anisotropy of the liquid crystal, polarisation microscopy studies were made using AxioImager A.1 polarising microscope (Carl Zeiss GmbH, Germany) equipped with a heatstage (Instec, USA). 
  Samples were prepared in commercial glass cells (E.H.C., Japan and WAT, Poland) with planar transparent indium tin oxide (ITO) electrodes (cell thickness: 2.5, 3, 5, 10, \qty{25}{\micro\meter}, ITO resistance: \qty{10}{\ohm}). 
  
  Cells with bare ITO/glass substrate were used as well as treated cells with alignment layers.  The aligning layers used are rubbed polyimide layers for LC planar alignment (parallel and antiparallel rubbing), and  cetyl-trimethyl-ammonium-bromide for the vertical alignment. The vertical alignment, however, could be achieved only in the non-polar nematic phase. Cells with interdigitated in-plane electrodes (IPS) were used for studying the in-plane switching (INSTEC, USA).
 
 The structure of the director field and the polarisation was investigated using polarising confocal laser scanning microscopy (Leica TCS SP8, CLSM). Generation of the optical second harmonic (SHG) was measured using the multiphoton laser of the confocal microscope. A tunable IR laser ($\lambda_{\rm{ex}}
 =\qty{880}{\nano\meter}$) was used as a fundamental light beam. The direction of maximal SHG efficiency was mapped in the SHG microscopy images to indicate the local direction of the polar axis. This is possible since the nonlinear coefficient $d_{33}\gg d_{31}$.
 
 The complex magnetic susceptibility was measured with commercially available AC susceptometer (Dynomag, RISE Acreo, Gothenburg, Sweden) with an excitation flux of \qty{0.5}{\milli\tesla} in a frequency range of \qty{1}{\hertz}
  to \qty{500}{\kilo\hertz}. The samples were measured in cylindrical vials containing \qty{100}{\micro\liter} of the hybrid-LC.

The optical transmission of the sample cells between crossed polarisers were carried out in a custom-made setup containing tungsten and laser sources, coil system, photodetector, CCD camera and a long-range microscope. The sample  was placed in oscillating magnetic field of a four-coil system, allowing arbitrary variation of the exciting field angle with respect to the position and/or the rubbing direction of the cell. Transmission was characterised using a cw He-Ne laser source $\lambda=\qty{632}{\nano\meter}$, $P=\qty{10}{\milli\watt}$ (JDS Uniphase) and the transmitted polarised light was detected by a photodetector (Thorlabs PDA36A2, detection range between 350–1100~nm). The measurement was performed in the frequency range of the magnetic field from \qty{0.1}{\hertz} to \qty{100}{\hertz}. The phase shift between the applied field and the transmission signal was determined from the analysis of the time series.  
 
 Magnetisation measurements were made using a fluxgate magnetometer sensor AS-UAP GEO-X Projekt Elektronik GmBH (Germany) and SQUID MPMS-3 (Quantum Design Europe). SQUID measurements were performed on samples filled in \qty{20}{mm} long thin capillaries with a diameter of \qty{1}{mm}. 
 
 The transient currents in magentic field across the LC cell were measured using a picoammeter (Keithley 6487) with an accuracy \qty{100}{\femto\ampere}. The picoammeter was connected to a digital oscilloscope PicoScope 6480.  
The cells were magnetised using a \qty{640}{\milli\tesla} electromagnet. The current transient was recorded on switching on the field (\qty{640}{\milli\tesla}) of the opposite polarity. 
In case of electric switching, we used cells with interdigitated comb electrodes (INSTEC) and recorded current transient across a \qty{4}{\kilo\ohm} resistor using an PicoScope 6400 digital oscilloscope.
 
\backmatter

\bmhead{Acknowledgments}

We acknowledge the financial support of Deutsche Forschungsgemeinschaft (Projects NA 1668/1-3,  LU 800/7-1, and ER 467/8-3), German Academic Exchange Service (DAAD), Programme for Project-Related Personal
Exchange Project Slovenia-Germany 57655627, and of the Slovenian Research and Innovation Agency through the bilateral project BI-DE/23-24-007 and the research core funding P2-0089 and P1-0192.  The study was also partly funded from the European Union's Horizon 2020 research and innovation programme by the project MAGNELIQ under grant agreement no 899285. The results reflect only the authors' view and the Commission is not responsible for any use that may be made of the information it contains.

\bigbreak
\pagebreak

\clearpage
\section*{Supplementary information}
\renewcommand{\thefigure}{S\arabic{figure}}
\setcounter{figure}{0}
\setcounter{section}{0}


\bigbreak
\subsection*{Supplementary Note 1. Abbreviations}
\begin{tabular}{l r}
   BaHF & barium hexaferrite  \\
   ITO & indium tin oxide \\
   LC  &  liquid crystal \\
   N & nematic phase\\
   N$_{\rm{F}}$ & ferroelectric nematic phase\\
   M & M-phase \\
   POM & polarising optical microscopy \\ 
   SHG & optical second harmonic generation \\
\end{tabular}

\subsection*{Supplementary Note 2. Electric and magnetic behaviour}
\begin{minipage}{\linewidth}
\makebox[\linewidth]{
\includegraphics[width=0.5\textwidth]{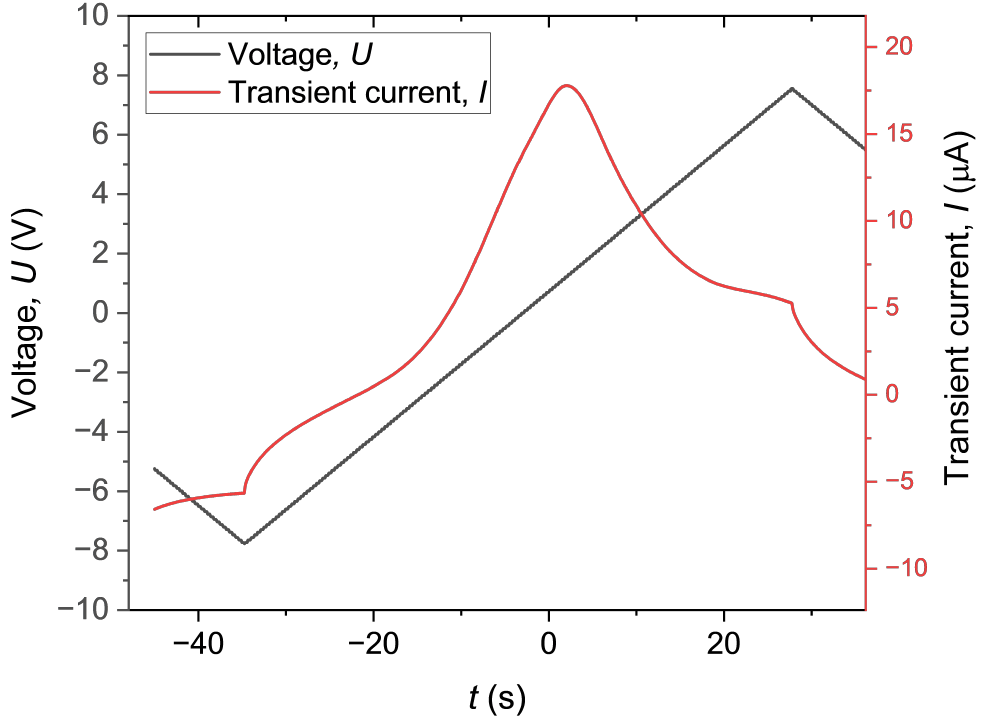}}
\captionof{figure}{\textbf{Current transient} in response to the triangular-wave voltage measured in a \qty{6}{\micro\meter} thin cell with polyimide aligning layers and in-plane electrodes ($T=\qty{21}{\celsius}$, inter-electrode distance \qty{20}{\micro\meter}).} \label{fig:current}
\end{minipage}

\begin{minipage}{\linewidth}
\makebox[\linewidth]{
\includegraphics[width=0.5\textwidth]{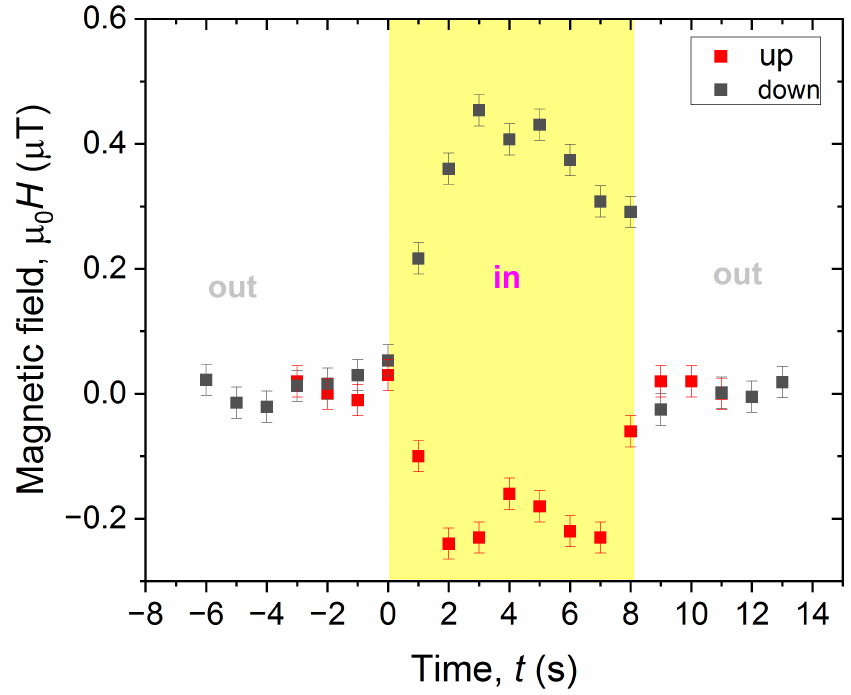}}
\captionof{figure}{\textbf{Remanent magnetisation in a thin cell:} Magnetic field measured at the surface of the \qty{10}{\micro\meter} thick V-type cell inserted under a fluxgate sensor. The cell was prepared by quenching in 700 mT field applied upwards (dataset "up") and downwards (dataset "down"). The measurements were done in a magnetically screened chamber.}\label{fig:SIb}
\end{minipage}

\bigbreak

\subsection*{Supplementary Note 3. Optical behaviour}
Polarising microscopy textures provide information on the molecular order and orientation in the liquid crystal phase. In uniformly aligned  thin nematic layers, the apparent colour is determined by the effective birefringence of the sample and the optical extinction direction is determined by the orientation of the optical axis.
Thick cells as in Fig~\ref{fig:SItextures}a show a disordered texture consisting of thin lines. The same texture remains in the M phase of the hybrid material (Fig~\ref{fig:SItextures}b). However, in the latter case, no reaction to applied electric or magnetic fields could be found. 

The optical transmittance of a thin layer uniformly planar-aligned nematics exhibits four extinctions and four maxima upon the \qty{360}{\degree} sample rotation. Twisted director structures can be determined through their optical rotation using slightly uncrossed polarisers or observing the textures in circularly polarised light.

The ferroelectric character of the molecular order in the N$_{\rm{F}}$-hybrid is manifested in the current transients in response to the applied triangular-wave voltage between the electrodes. The current peak is produced by the polarisation reversal and is similar to that measured in the N$_{\rm{F}}$ of the pure compound (Fig.~\ref{fig:current}).

In thin (\qty{2.5}{\micro\meter}) cells with antiparallel rubbed polyimide aligning layers, the nematic director favours twisted configuration in the N$_{\rm{F}}$ phase. However, occasionally, uniformly aligned domains occur too. In case of the hybrid N$_{\rm{F}}$, stripped domains (Type 1) with the preferred orientation along the director are observed as well as the twisted domains (Type 2) (Fig~\ref{fig:SItextures2}a,b). The domains can also be distinguished by their optical transmittance as a function of the angle $\varphi$ between the polariser and the cell's aligning direction. Compared to Type 1 domains, the Type 2 domains show much smaller difference between the transmission maxima and minima (Fig~\ref{fig:SItextures2}c).

At a high magnification, the striped domains additionally display a splay deformation of the director. This can be seen from the displacement of the extinction brushes upon rotating the sample between crossed polarisers (Fig~\ref{fig:SItextures}a-c).


\begin{minipage}{\linewidth}
\makebox[\linewidth]{
\includegraphics[width=\textwidth]{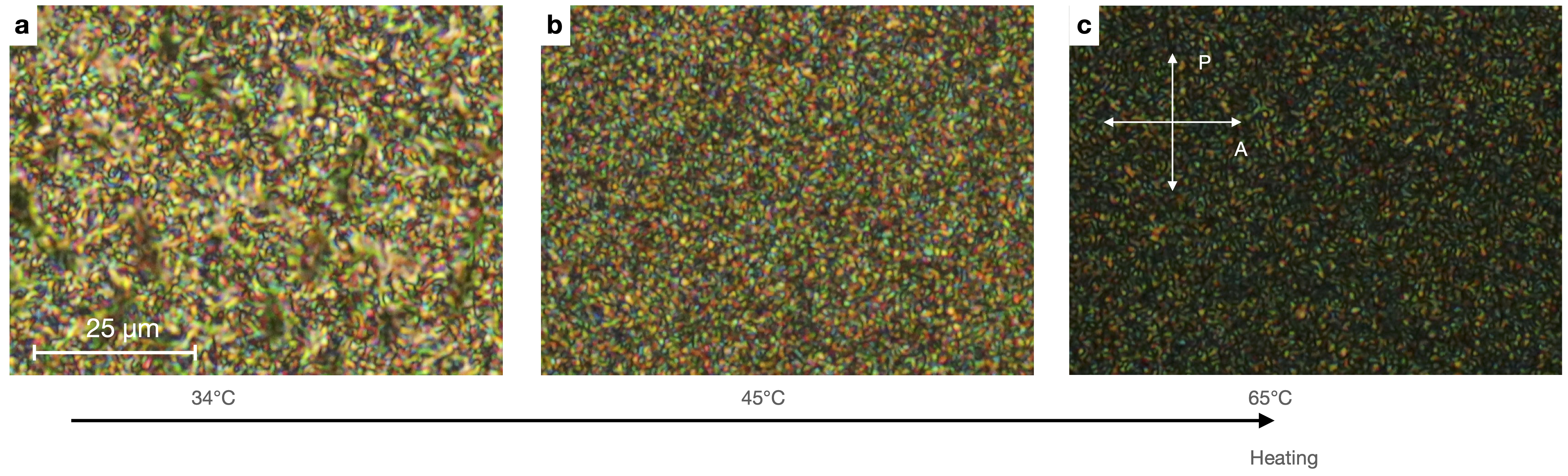}}
\captionof{figure}{\textbf{Temperature-driven transformations of polarising microscopy textures in hybrid-LC:} on heating in \qty{10}{\micro\meter} cell with polyimide rubbed substrates in the hybrid N$_{\rm{F}}$ phase at $T=\qty{34}{\celsius}$ \textbf{a}, \textbf{b} M phase $T=\qty{45}{\celsius}$, and \textbf{c} in the N phase $T=\qty{65}{\celsius}$.} \label{fig:SItextures}
\end{minipage}
\bigbreak

\begin{minipage}{\linewidth}
\makebox[\linewidth]{
\includegraphics[width=0.8\textwidth]{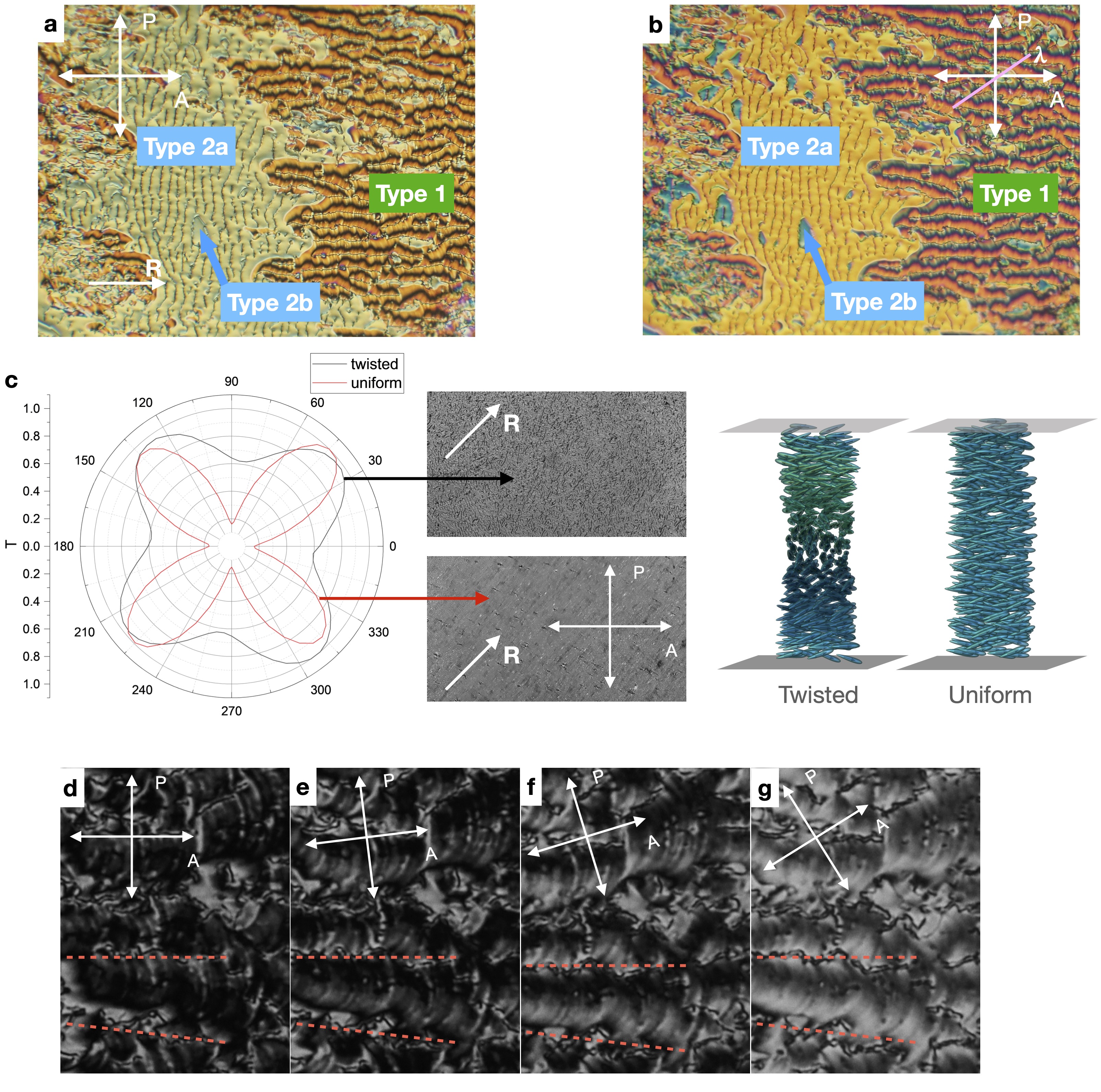}}
\captionof{figure}{\textbf{Optical domain structures in thin cells:}  
	\textbf{a} POM image of the N$_{\rm{F}}$ texture in a \qty{2.5}{\micro\meter} P-type cell between crossed polarisers and 
	\textbf{b} with a full-wave plate inserted.
	\textbf{c} Normalised transmittance in the twisted and uniform stripped domains as a function of the angle between the polariser and the cell's aligning layer direction $\mathbf{R}$ with corresponding optical textures. On the right side displays schematics of the twisted and uniformly aligned nematic structures, corresponsingly.
	\textbf{d - g} Director stripes in Type 1 domains depicted at different angles between the aligning layer direction an the polarisers. The shift of the extinction brush suggests a splay deformation of the director within the stripe.
	} \label{fig:SItextures2}
\end{minipage}
\bigbreak

\subsection*{Supplementary Note 4. Morphology of nanoplatelet/N$_{\rm{F}}$ hybrid confined to 10~\textmu m V-cell.}
In the V-cells, the aligning layer favouring orthogonal alignment of conventional nematic phases is not sufficient to achieve the orthogonal alignment in the N$_{\rm{F}}$ phase. For the morphological studies isotropic 0.3~wt$\%$  nanoplatelet/N$_{\rm{F}}$ hybrid-LC was filled into V-cells and quenched to the N$_{\rm{F}}$ phase in \qty{0.5}{\tesla} homogenous magnetic field perpendicular to the substrate surface. The nanoplatelet/N$_{\rm{F}}$ hybrids exhibit even more complex and various morphologies resulted from the competition between the aligning interactions, bulk elasticity and the distortions introduced by the nanoparticles. Large disordered (Fig.~\ref{fig:SIp1}a), grainy (Fig.~\ref{fig:SIp1}b), striped (Fig.~\ref{fig:SIp1}c) and uniform domains (Fig.~\ref{fig:SIp1}d,e) were observed. Similarly to the P-cells, the grainy texture consists of disclination lines which are responsive to magnetic fields in the range of \qty{100}{\milli\tesla}. Some domains, however, did not show any magnetic response as in (Fig.~\ref{fig:SIp1}a, b) suggesting assembly of the nanoplatelets in the non-magnetic (antiparallel) fashion. 

Magnetically responsive regions with uniform patches on a \qty{20}{\micro\meter} scale were formed typically outside the ITO-electrode area of the cell. Such regions were not observed in P-cells.
In these disclination-free areas of Fig.~\ref{fig:SIp1}d repeated application of a horizontal magnetic field of opposite signs alters the transmitted light intensity ($I$) and shifts the domain borders (Fig.~\ref{fig:SIp1}e). After the removal of the magnetic field, the texture relaxes to a state different from the initial one.
To explore the effect of the horizontal magnetic field, $I$ was measured in several regions of different domains at different orientation of the crossed polarisers. Fig.~\ref{fig:SIp2} shows the intensity profiles with and without magnetic field. The results show that without magnetic field (Fig.~\ref{fig:SIp2}a) director $\mathbf{n}$ varies domain by domain. However, application of the magnetic field (Fig.~\ref{fig:SIp2}b) has an ordering effect, since most of the intensity profiles almost coincide with intensity maxima at \qty{0}{\degree} and \qty{90}{\degree}, i.e. when the external field is aligned along one of the polarisers.

\begin{minipage}{\linewidth}
\makebox[\linewidth]{
\includegraphics[width=0.8\textwidth]{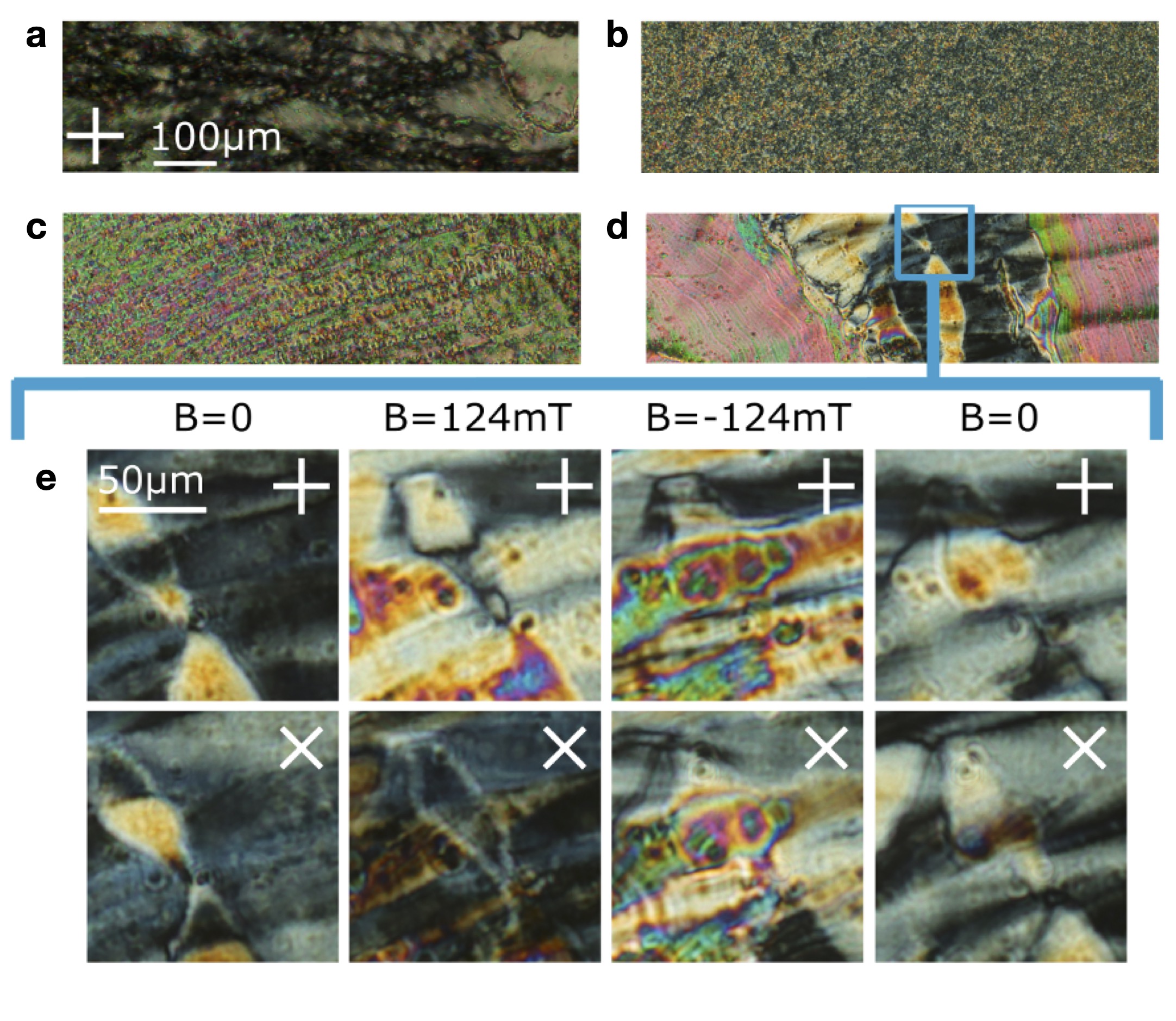}}
\captionof{figure}{\textbf{Morphological observations of nanoplatelet/N$_{\rm{F}}$ hybrid confined to  10~\textmu m V-cell using POM.} 
	\textbf{a} Disordered magnetically irresponsive texture.
	\textbf{b} Magnetically irresponsive grainy texture.
	\textbf{c} Magnetically responsive stripped texture.
	\textbf{d} Magnetically responsive uniform texture.
	\textbf{e} Magnetooptical response in the region marked by a blue rectangle in \textbf{d} The images are taken with crossed polarisers at the magnetic field strengths marked above images. In the upper row, the polariser is  aligned horizontally, and in the bottom row, it is rotated by \qty{45}{\degree} (white marks show the polariser's alignment). 
	}\label{fig:SIp1}
\end{minipage}

\begin{minipage}{\linewidth}
\makebox[\linewidth]{
\includegraphics[width=0.9\textwidth]{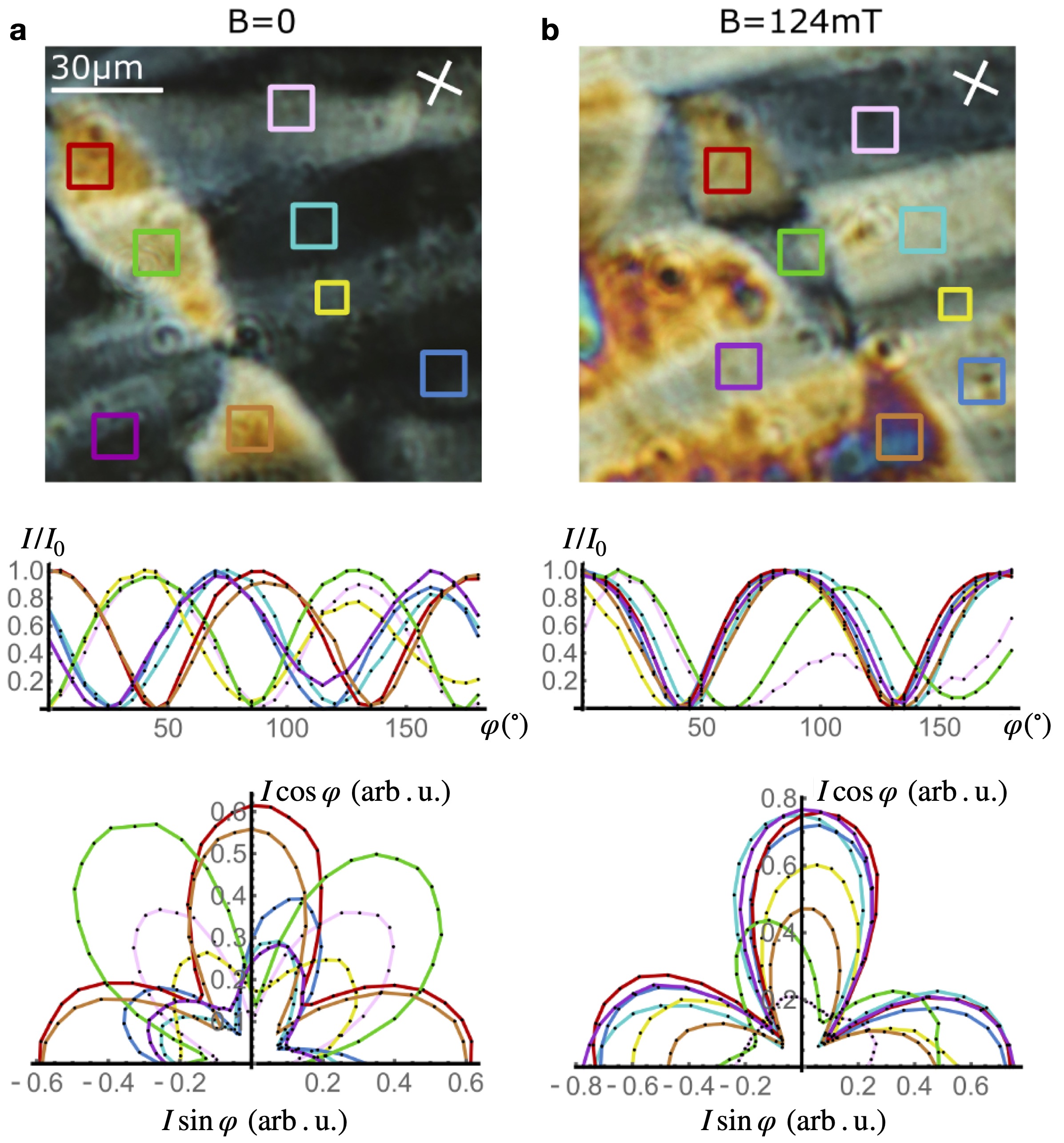}}
\captionof{figure}{\textbf{Polarising microscopy analysis of the polydomain structure of nanoplatelet/N$_{\rm{F}}$ hybrid confined to a 10~\textmu m  V-cell.} 
Results shown in Figure~\ref{fig:SIp1}d are here further analysed for the cases \textbf{a} without external magnetic field  and \textbf{b} when the in-plane magnetic field  was applied. Eight different domains were analysed each denoted by a distinct color. The plots below the microscopic images show the normalised intensity of transmitted light depending on the orientation of the polarisers. The bottom parametric plots display the non-normalised intensity of transmitted light (average value of the red, blue, and green pixels in the photograph) for different angles. Photographs were taken as the sample was rotated with respect to the crossed polarisers. The external magnetic field was always applied in the same direction with respect to the sample. Images \textbf{a} and \textbf{b} were taken for the sample aligned at \qty{115}{\degree}. The borders between different domains were clearly visible due to significant changes in the transmitted light intensity. 
}\label{fig:SIp2}
\end{minipage}

\subsection*{Supplementary Note 5. Optical response of Type 1 domain of nanoplatelet/N$_{\rm{F}}$ hybrid confined to 2.5~\textmu m P-cell to external magnetic field.}
In the Type 1 domains of nanoplatelet/N$_{\rm{F}}$ hybrid confined to \qty{2.5}{\micro\meter} P-cell the external magnetic field disturbs the disclination lines and causes buckling instability as observed by POM in Figure \ref{fig:schemes}a. The disclination lines are coupled to the electric polarisation orientation of the surrounding N$_{\rm{F}}$ phase, so their magnetically induced movement causes rearrangement of the polarisation structure within the sample. This is shown in Figure~\ref{fig:SIp3} where SHG signal is recorded in dependence on the polarisation orientation of the incoming probing IR beam. Second order dielectric susceptibility tensor $d_{ij}$ for this class of compounds satisfies the conditions for its components $d_{33}\>>d_{31}$, where the 33 axis is along the nematic director. This allows determining  the local polar axis alignment from the direction of maximal SHG efficiency. (see 
 Folcia CL, Ortega J, Vidal R, Sierra T, Etxebarria J. The ferroelectric nematic phase: an optimum liquid crystal candidate for nonlinear optics. Liq Cryst. 2022:1-8. doi:10.1080/02678292.2022.2056927 and 
    Sebasti\'an N, Lov\v{s}in M, Berteloot B, et al. Polarization patterning in ferroelectric nematic liquids via flexoelectric coupling. Nat Commun. 2023;14(1):3029. doi:10.1038/s41467-023-38749-2
  )
\begin{minipage}{\linewidth}
\makebox[\linewidth]{
\includegraphics[width=0.9\textwidth]{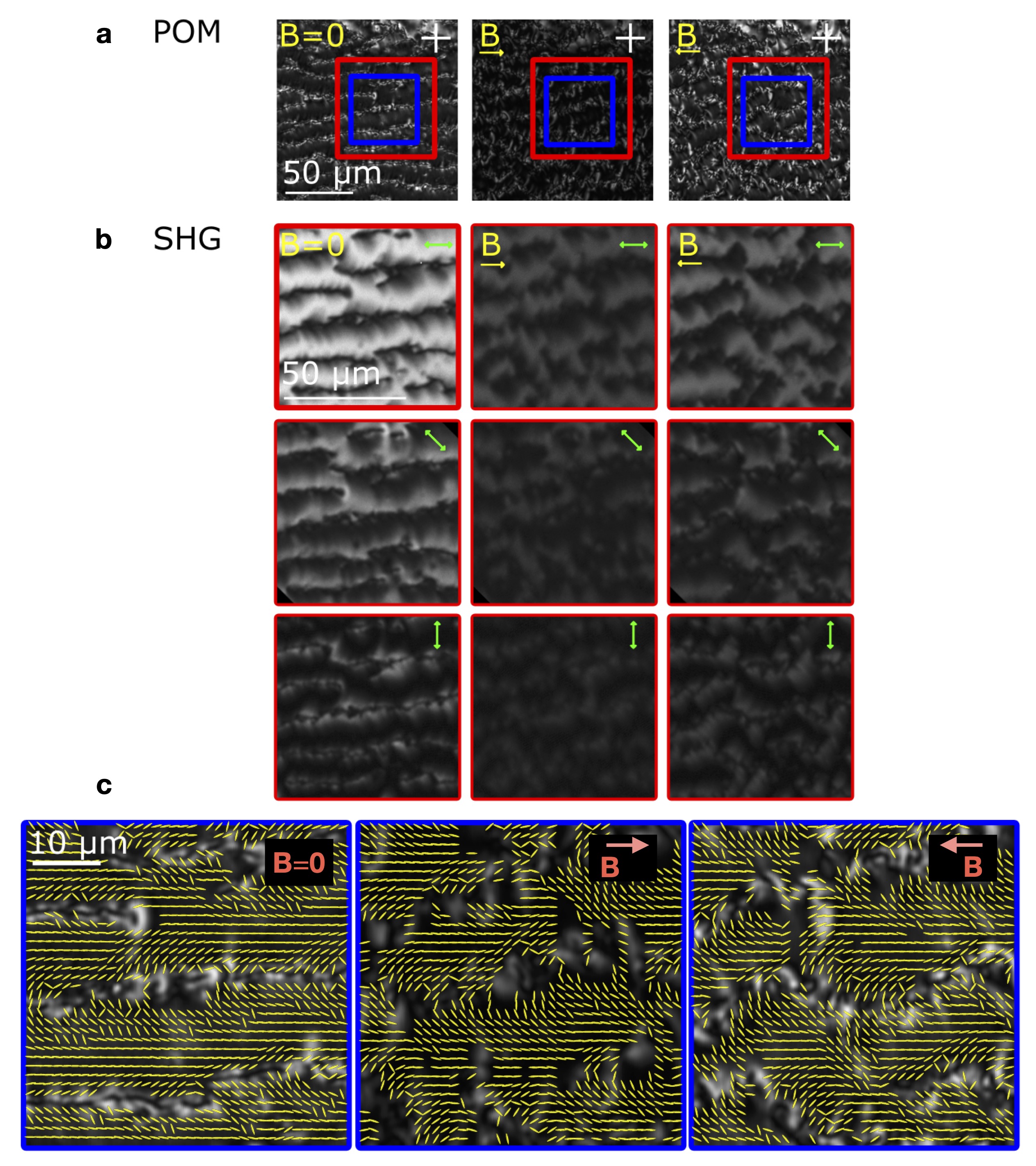}}
\captionof{figure}{\textbf{POM and SHG analysis of Type 1 domains of nanoplatelet/N$_{\rm{F}}$ hybrid confined to 2.5~\textmu m P-cell}. \textbf{a} POM images correspond to the cases without magnetic field and with applied magnetic fields of opposite polarities. 
Part \textbf{b} shows the SHG microscopy of the corresponding regions in \textbf{a}  marked by red rectangles. The images were taken at three different orientations of polarisation of the incoming IR beam (green arrows) without an analyser. The figures show how the external magnetic field affects the SHG efficiency. Analysis of the SHG microscopy images allows us to extract the direction of the highest efficiency corresponding to the direction of the local polarisation. In \textbf{c} the polarisation field marked in yellow is overlain with the POM images marked blue in \textbf{a}. }\label{fig:SIp3}
 \end{minipage}

\subsection*{Supplementary movie 1}
Bistable magnetooptical response of a network texture in a \qty{6}{\micro\meter} thick cell treated for vertical alignment. The image width is \qty{6}{\micro\meter}, crossed polarisers are aligned along the image sides.  
\subsection*{Supplementary movie 2}
Rearrangement of the disclination network in response to an external magnetic field observed in a \qty{5}{\micro\meter} cell with a bare glass substrate (no aligning treatment).  The image width is \qty{250}{\micro\meter}, crossed polarisers are aligned along the image sides.
\subsection*{Supplementary movie 3}
SHG microscopy on a \qty{5}{\micro\meter} cell exposed to an external magnetic field for Fig.~\ref{fig:textures2}f. 

\subsection*{Supplementary movie 4}
Field-induced rearrangement of the disclination lines in striped domains aligned along the alignment direction a \qty{3}{\micro\meter} cell with a planar alignment. The image width is \qty{190}{\micro\meter}, crossed polarisers are aligned along the image sides.  

\subsection*{Supplementary movie 5}
Soliton-like distortion propagation along the disclination lines in the Type 1 domains  observed in a \qty{2}{\micro\meter} thin cell with antiparallel planar surface treatment. The image width is \qty{205}{\micro\meter}, crossed polarisers are aligned along the image sides. 

\subsection*{Supplementary movie 6}
Switching in twisted domains in a 3 µm thick cell with polyimide rubbing ($c=0.2$~wt$\%$). The image width is \qty{225}{\micro\meter}, crossed polarisers are aligned along the sides of the image.
\subsection*{Supplementary movie 7}
Propagation of the director distortion along a disclination line after poling the magnetic field ($B=\qty{140}{\milli\tesla}$, the image height is \qty{12}{\micro\meter}, cell thickness $d=\qty{2.5}{\micro\meter}$

\clearpage
\end{document}